\documentstyle[aps,psfig,floats]{revtex}

\begin{document}
\draft

\title{Coherent structures in a turbulent environment}
\author{F.\ Spineanu$^{1,2}$, M. Vlad$^{1,2}$ \\
$^1${\small Association Euratom-C.E.A. sur la Fusion, C.E.A.-Cadarache, }\\
{\small \ \ F-13108 Saint-Paul-lez-Durance, France}\\
{\small \ }$^2${\small National Institute for Laser, Plasma and Radiation
Physics, }\\
{\small \ \ \ P.O.Box MG-36, Magurele, Bucharest, Romania\ \smallskip }}
\maketitle

\begin{abstract}
A systematic method is proposed for the determination of the statistical
properties of a field consisting of a coherent structure interacting with
turbulent linear waves. The explicit expression of the generating functional
of the correlations is obtained, performing the functional integration on a
neighbourhood in the function space around the soliton. The results show
that the non-gaussian fluctuations observed in the plasma edge can be
explained by the intermittent formation of nonlinear coherent structures.
\end{abstract}


\section{Introduction}

In a recent work \cite{PRLflmadi} it has been proposed a systematic
analytical method for the investigation of the statistical properties of a
coherent structure interacting with turbulent field. The method is here
developed in detail and new possible applications or developments arise.

The nonlinearity of the dynamical equations of fluids and plasma is the
determining factor in the behaviour of these systems. The current
manifestation is the generation, from almost all initial conditions, of
turbulent states, with an irregular aspect of fluctuations implying a wide
range of space and time scales. The fluctuations seem to be random and a
statistical characterization of the fluctuating fields is appropriate.
However it is known both from theory and experiment that the same fields can
have, in particular situations, stable and regular forms which can be
identified as coherent structures, for example solitons and vortices. For
most general conditions one should expect that these aspects are both
present and this requires to study the mixed state consisting of coherent
structures and homogeneous turbulence.

Numerical simulations of magnetohydrodynamics show that in general cases a
coherent structure emerges in a turbulent plasma, it moves while deforming
due to the interactions with the random fields around it and eventually is
destroyed. In plasma turbulence a coherent structure is build up by the
inverse spectral cascade or by merging and coalescence of small-scale
structures \cite{Montgomery}, \cite{Biskamp}, \cite{Craddock}. The
nonlinearity of the equations for the drift waves in a non-uniform,
magnetized plasma permits the formation of solitary waves in addition to the
usual small-amplitude dispersive modes. The convective nonlinearity (of the
Poisson bracket type), can lead to {\it low-frequency convective structures}
in magnetized plasma \cite{Su}, \cite{KonoMyia}, \cite{Tajima1}, \cite
{Hazeltine}, \cite{Nycander}. The structures are not solitons in the strict
sense but are very robust. It is even possible that the state of plasma
turbulence can be represented as a superposition of coherent vortex
structures (generated by a self-organization process) and weakly correlated
turbulent fluctuations.

Naturally, the coherent structures influences the statistical properties of
the fields (the correlations), in particular the spectrum. In this context
it is usual to say that the deviation of the correlations of the fluctuating
fields from the gaussian statistics is associated with the presence of the
coherent structures and it is named {\it intermittency}. Numerical
simulations \cite{Kinney} of the $2$-dimensional Navier-Stokes fluid
turbulence have shown coherent structures evolving from random initial
conditions and in general energy spectra steeper than $k^{-3}$ have been
atributed to intermittency (patchy, spatial intermittent paterns). These
coherent structures are long-lived and disappear only by coalescence, the
latter being manifested as spatial intermittency. In these studies it was
underlined that the coherent structures have effects which cannot be
predicted by closure methods applied to mode-coupling hyerarchies of
equations.

The difficulty of the analytical description consists in the absence from
theory of well established technical methods to investigate the plasma
turbulence in the presence of coherent structures. While for the
instability-induced turbulence (via nonlinear mode-coupling) systematic
renormalization procedures have been developed, the problem of the
simultaneous presence of coherent structures and drift turbulence has not
received a comparable detailed description.

In the recently proposed method \cite{PRLflmadi}, the starting point is the
observation that the coherent structure and the drift waves, although very
different in form, are similar from a particular point of view: the first
realizes the extremum and the later is very close to the extremum of the
action functional that describes the evolution of the plasma. The analytical
framework is developed such as to exploit this feature and is based on
results from well established theories: the functional statistical study of
the properties of the classical stochastic dynamical systems (in the
Martin-Siggia-Rose approach); the perturbed Inverse Scattering Transform
method, allowing to calculate the field of perturbed nonlinear coherent
structures; the semi-classical approximation in the study of the quantum
particle motion in multiple minima potentials.

The dilute gas of plasma solitons has been studied by Meiss and Horton \cite
{Pf25-82-1838} who assumed a probability density function of the amplitudes
characteristic of the Gibbs ensemble. We analyse the same nonlinear equation
but take into account the drift wave turbulence.

A brief discussion on the closure methods developed in the study of drift
wave turbulence provides us the argumentation for the need of a different
approach (Section 2). The Section 3 contains a description of the general
lines of the method proposed. A more technical presentation of the
calculation is given in the subsection 3.2. The particular case of the drift
wave equation is developed in detail in the Section 4 and in the Section 5
the explicit expression of the generating functional is used for the
calculation of the correlation functions. The results and the conclusions
are presented in the last Section. Some details of calculations are given in
the Appendix.

\section{The nonlinear dynamical equations}

We consider the plasma confined in a strong magnetic field and the drift
wave electric potential in the transversal plane $\left( x,y\right) $ where $%
y$ corresponds to the poloidal direction and $x$ to the radial one in a
tokamak. We shall work with the radially symmetric Flierl-Petviashvili
soliton equation \cite{csf9-98-2007} studied in Ref.\cite{Pf25-82-1838} : 
\begin{equation}
\left( 1-\rho _{s}^{2}\nabla _{\perp }^{2}\right) \frac{\partial \varphi }{%
\partial t}+v_{d}\frac{\partial \varphi }{\partial y}-v_{d}\varphi \frac{%
\partial \varphi }{\partial y}=0  \label{eqh1}
\end{equation}
where $\rho _{s}=c_{s}/\Omega _{i}$, $c_{s}=\left( T_{e}/m_{i}\right) ^{1/2}$
and the potential is scaled as $\varphi =\frac{L_{n}}{L_{T_{e}}}\,\frac{%
e\Phi }{T_{e}}$ . Here $L_{n}$ and $L_{T}$ are respectively the gradient
lengths of the density and temperature. The velocity is the diamagnetic
velocity $v_{d}=\frac{\rho _{s}c_{s}}{L_{n}}$. The condition for the
validity of this equation is: $\left( k_{x}\rho _{s}\right) \left( k\rho
_{s}\right) ^{2}\ll \eta _{e}\frac{\rho _{s}}{L_{n}}$, where $\eta _{e}=%
\frac{L_{n}}{L_{T_{e}}}$.

The exact solution of the equation is 
\begin{equation}
\varphi _{s}\left( y,t;y_{0},u\right) =-3\left( \frac{u}{v_{d}}-1\right)
\sec h^{2}\left[ \frac{1}{2\rho _{s}}\left( 1-\frac{v_{d}}{u}\right)
^{1/2}\,\left( y-y_{0}-ut\right) \right]  \label{eqh7}
\end{equation}
where the velocity is restricted to the intervals $u>v_{d}$\ \ \ or\ \ \ $%
u<0 $. The function is represented in Fig.1. In the Ref. \cite{Pf25-82-1838}
the radial extension of the solution is estimated as: $\left( \Delta
x\right) ^{2}\sim \rho _{s}L_{n}$. In our work we shall assume that $u$ is
very close to $v_{d}$ , $u\gtrsim v_{d}$ (i.e. the solitons have small
amplitudes).

The nonlinear equations for the drift waves are known to generate as
solutions irregular turbulent fields but also exact coherent structures of
the type (\ref{eqh7}), depending on the initial conditions. Typical
statistical quantities are the correlations, like: $\left\langle \varphi
\left( x,y,t\right) \varphi \left( x^{\prime },y^{\prime },t^{\prime
}\right) \right\rangle \sim \left| {\bf x-x}^{\prime }\right| ^{\zeta
}\left| t-t^{\prime }\right| ^{z}$ , where for the homogeneous turbulence
the exponents $\zeta $ and $z$ are calculated by the theory of
renormalization or by spectral balance equations, using closure methods ( 
\cite{Diamond}). Various closure methods have been developed as
perturbations around gaussianity and they are valid for small deviation from
the gaussian statistics \cite{Krommes}. We see intuitively that this
approach cannot be extended to the description of the coherent structures.
This can also be seen in more analytical terms. A quantity which is
unavoidable in the calculation of the correlations is the average of the
exponential of a functional of the fluctuating field, consider simply $%
\left\langle \exp \left( \varphi \right) \right\rangle $ (for example in the
inverse of the Vlasov operator, using the Fourier transformation, the
potential appears in the formal expression of the trajectory, i.e. at the
exponent). This quantity can be written schematically as \cite{Krommes}: 
\begin{equation}
\left\langle \exp \left( \varphi \right) \right\rangle =\exp \left[ \sum_{n}%
\frac{1}{n!}\left\langle \left\langle \varphi ^{n}\right\rangle
\right\rangle \right]  \label{cumulant}
\end{equation}
where $\left\langle \left\langle \varphi ^{n}\right\rangle \right\rangle $
represents the cumulant of order $n$ (i.e. the irreducible part of the
correlation, after substracting the combinations of the lower order
cumulants). For a gaussian statistics the first two cumulants are different
of zero ($n=1$ : average and $n=2$ : dispersion), all others are zero.
Non-vanishing of the higher order cumulants is the signature of non-gaussian
statistics. In the perturbative renormalization we assume slight deviation
from gaussianity, i.e. small absolute values of the next order cumulants
(e.g. the kurtosis must be close to $3$, the gaussian value) and vanishing
of the higher order cumulants. This assumption is obviously invalid in the
case of coherent structures. The field of a coherent structure has long
range, persistent correlations imposed by its regular geometry, which
naturally requires non-vanishing very large order cumulants ({\it i.e.} many
terms in the sum at the exponent in Eq.(\ref{cumulant})) and excludes any
perturbative expansion.

In particular, the closure of the nonlinear equation for the two-point
correlation (based on the retaining the directly interacting triplet) can
account for the small scale correlations related to the space-dependent
relative diffusion, i.e. the clump effect (\cite{Misguich}, \cite{Balescu}, 
\cite{Diamond}), but the spectrum obtained in this framework cannot account
for the possible existence of the coherent structures. This clearly suggests
that we must find a different approach.

\section{Coherent structures in a turbulent background}

\subsection{The outline of the method}

We present the basic lines of an approach which can provide a statistical
description of the coherent structure in a turbulent background. The
physical origin of this approach is the observation that the non-linear
equation whose solution is the coherent structure (the vortex soliton) also
has classical drift waves as solutions, in the case of very weak
nonlinearity. In a certain sense (which will become more clear further on),
the vortex soliton and the drift waves belong to the same family of
dynamical configurations of the plasma. Our approach, which is designed to
put in evidence and to exploit this property, consists of the following
steps.

We start by constructing the action functional of the system. The dynamical
equation is the Euler-Lagrange equation derived from the condition of
extremum of this functional and the exact solution is the vortex soliton (%
\ref{eqh7}).

By using the exponential of the action we construct the generating
functional of the irreducible correlations of $\varphi $. This functional
contains all the information on the coherent structure and the drift
turbulence. The correlations are obtained via functional differentiations.
This requires the formal introduction of a perturbation of the system,
through the interaction with an external current. Throughout the work, this
perturbation will be considered a small quantity and finally it will be
taken zero.

The generating functional is by definition a functional integral over all
possible configurations of the system and this integral must be calculated
explicitely. The simplest thing to do is to determine the configuration of
the system (with space and time dependence) which extremises the action, by
equating the first functional variation of the action with zero and solving
this equation: this will give the vortex soliton (modified due to the small
interaction term). Then one should replace this solution in the expression
of the action. This is the lowest approximation and it does not contain
anything related to the drift wave trubulence.

At this point we can benefit of the particular physics of the drift waves.
The vortex soliton is the exact solution of the fully nonlinear equation and
is a localized potential perturbation with regular, cylindrical symmetric
form. The linear drift waves are harmonic potential perturbations
propagating with constant velocity (the diamagnetic velocity in the case of
the drift poloidal propagation in tokamak). Although the drift waves have
very different geometry they are solutions of the same equation as the
vortex, but for negligible magnitude of the nonlinear term. The drift waves
do not exactly realize the extremum of the action functional, but obtain an
action very close to this extremum. This means that the drift waves and the
vortex soliton are close in the function space in the sense of the measure
defined by the exponential of the action. In other terms the drift waves are
in a functional neighbourhood of the vortex (for this measure). This
suggests to perform the functional integral with better approximation, which
means to perform the integration over a functional neighbourhood of the
vortex solution. This will automatically include the drift waves in the
generating functional of correlation which so will contain information on
both the coherent structure and the drift waves. The function space
neighbourhood over which the functional integration is extended is limited
by the measure (exponential of the action) which severly penalizes all
configurations of the system which are far from the solution realizing the
extremum (i.e. the vortex soliton). As in any stationary phase method there
are oscillations which strongly suppress the contribution of the
configurations which are far (in the sense of the measure) from the soliton.
In practice we shall expand the action in a functional Taylor series around
the soliton solution and keep the term with the second functional derivative.

In this perspective the drift waves appear as fluctuations around the
soliton solution. This is compatible with the numerical simulations which
show that the vortices are accompagned by a tail of drift waves. During the
interaction of the vortices linear drift waves are ``radiated'' \cite
{Horton2}. On the other hand, the analytical treatment of the perturbed
vortex solution by the perturbed Inverse Scattering Transform shows similar
tail of perturbed field, following the soliton. This strengthens our
argument that integrating close to the vortex means to include the drift
waves in the generating functional.

The functional integral can be performed exactly and we determine the
generating functional of the potential correlations. We shall calculate the
two-point correlation by performing double functional derivative at the
external current.

\subsection{Expansion around a soliton}

\subsubsection{The action and the generating functional of the correlations}

The analytical framework is similar to the model of quantum fluctuations
around the instanton solution in the semi-classical calculation of the
transition amplitude for the particle in a two-well potential (see reference 
\cite{qcd}). Let us write formally the equation for a nonlinear plasma waves
as 
\begin{equation}
\widehat{{\it O}}\varphi =0  \label{qcd1}
\end{equation}
where the field $\varphi \left( x,y,t\right) $ represents the ``field''
(coherent structure + drift waves) and the operator $\widehat{{\it O}}$ is
the nonlinear operator of the equation (\ref{eqh1}). This equation should be
derived from the condition of extremum of an action functional which must
reflect the statistical nature of our problem. The field $\varphi $ obeys a
purely deterministic equation, but the randomness of the initial conditions
generates a statistical ensemble of realizations of the system evolutions
(space-time configurations). We shall follow the Martin-Siggia-Rose method
of constructing the action functional but in the path-integral formalism,
for which we give in the following a very short description (\cite{MSR}, 
\cite{Jensen}, \cite{florinmadi1}). First, we consider a formal extension
from the statistical ensemble of realizations of the system's space-time
configurations to a larger space of functions which may include even
non-physical configurations. Every function is discretized in space and
time, so it will be represented as a collection of varables $\varphi _{i}$ ,
each attached to the corresponding space-time point $i$. In this space of
functions, the selection of the configurations which correspond to the
physical ones (solutions of the equation of motion) is performed through the
identification with Dirac delta-functions, in every space-time point 
\begin{equation}
\prod_{i}\delta \left[ \varphi _{i}-\varphi \left( x_{i},y_{i},t_{i}\right) %
\right] \delta \left[ \widehat{{\it O}}\varphi \right]  \label{qcd2}
\end{equation}
and integration over all possible functions $\varphi ,{\it \ }${\it i.e.}
over the ensemble of independent variables $\varphi _{i}$. Using the Fourier
representation for every $\delta $ function we get 
\begin{equation}
\int \prod_{i}d\varphi _{i}\int \prod_{i}d\chi _{i}\;\exp \left[ i\chi _{i}%
\widehat{{\it O}}\varphi \left( x_{i},y_{i},t_{i}\right) \right]
\label{qcd3}
\end{equation}
Going to the continuum limit, a new function appears, $\chi \left(
x,y,t\right) $ which is similar to the Fourier conjugate of $\varphi $. The
generating functional of the correlation functions is 
\begin{equation}
{\it Z}=\int D\left[ \varphi \left( {\bf x,}t\right) \right] \,D\left[ \chi
\left( {\bf x,}t\right) \right] \exp \left\{ i\int d{\bf x}^{\prime
}dt^{\prime }\chi \left( {\bf x}^{\prime },t^{\prime }\right) \widehat{{\it O%
}}\varphi \left( {\bf x}^{\prime },t^{\prime }\right) \right\}  \label{qcd4}
\end{equation}
where the functional measures have been introduced and ${\bf x\equiv }\left(
x,y\right) $.

The random initial conditions $\varphi _{0}\left( y\right) $ can be included
by a Dirac $\delta $ functional: $\delta \left( \varphi \left(
t_{0},y\right) -\varphi _{0}\left( y\right) \right) $. As explained in \cite
{PRLflmadi}, instead of this exact treatment (accessible only numerically)
we exploit the particularity of our approach, {\em i.e.} the connection
between the functional integration and the delimitation of the statistical
ensemble: the way we perform the functional integration is an implicit
choice of the statistical ensemble. We choose to build implicitely the
statistical ensemble, collecting all configurations which have the same type
of deformations (given in our formulas by $\widetilde{\chi }_{J}$). All
these configurations belong to the neighbourhood of the extremum in function
space and we take them into account, by performing the integration over this
space. In doing so we assume that the ensemble of perturbed configurations
induced by an ``external'' excitation ($J$ below) of the system is the same
as the statistical ensemble of the system's configurations evolving from
random initial conditions.

We must add to the expression in the integrand at the exponential a linear
combination related to the interaction of the fields $\varphi $ and $\chi $
with external currents $J_\varphi $ and $J_\chi $: 
\begin{eqnarray}
{\it Z} &\rightarrow &{\it Z}_J=\int D\left[ \varphi \left( {\bf x,}t\right) %
\right] \,D\left[ \chi \left( {\bf x,}t\right) \right] \exp \left\{
iS_J\right\}  \label{qcd5} \\
S_J &\equiv &\int d{\bf x}^{\prime }dt^{\prime }\left[ \chi \left( {\bf x}%
^{\prime },t^{\prime }\right) \widehat{{\it O}}\varphi \left( {\bf x}%
^{\prime },t^{\prime }\right) +J_\varphi \varphi +J_\chi \chi \right] 
\nonumber
\end{eqnarray}
It is now possible to obtain correlations by functional differentiation, for
example 
\begin{equation}
\left\langle \varphi \left( x_2,y_2,t_2\right) \varphi \left(
x_1,y_1,t_1\right) \right\rangle =\left. \frac 1{{\it Z}_J}\frac{\delta ^2%
{\it Z}_J}{\delta J_\varphi \left( x_2,y_2,t_2\right) \delta J_\varphi
\left( x_1,y_1,t_1\right) }\right| _{{\bf J}=0}  \label{deffifi}
\end{equation}

For the explicit calculation of the generating functional we need the
functions $\varphi $ and $\chi $ which extremize the action 
\begin{eqnarray}
\frac{\delta S_J}{\delta \varphi } &=&0  \label{deltas} \\
\frac{\delta S_J}{\delta \chi } &=&0  \nonumber
\end{eqnarray}

\subsubsection{Schema of calculation of the generating functional}

In the absence of the current ${\bf J}$ the equations (\ref{deltas}) have as
solutions for $\varphi $ the nonlinear solitons (vortices) \cite{Eilenberger}%
, \cite{DrazinJohnson}. More generally, the basic solution of the KdV
equation (on which the Flierl-Petviashvili equation can be mapped) is the
periodic cnoidal function which becomes, when the modulus of the elliptic
function is close to $1$, the soliton. When the distance between the centres
of the solitons is much larger than their spatial extension (dilute gas) the
general solution can be written as a superposition of individual solitons,
with different velocities and different positions \cite{Pf25-82-1838}. For
simplicity we shall consider in this work a single vortex soliton and in the
last Section we shall comment on the extension of the method to many
solitons.

The position of the centre of the soliton rises the difficult problem of the 
{\it zero modes }\cite{qcd}. Except for a brief comment about the relation
of the zero modes with the gaussian functional integration (see below), we
shall avoid this problem and postpone the discussion of this topic to a
future work.

In the presence of the external current ${\bf J}$, the equations resulting
from the extremization of the action $S_J$ become inhomogeneous, and the
solutions are {\it perturbed }solitons. This point is technically
non-trivial and we shall use the results obtained by Karpman \cite{Karpman}
who considered the Inverse Scattering Transform method applied to the
perturbed soliton equation. We find the approximate solution $\varphi _{Js}$
and $\chi _{Js}$ of the inhomogeneous equations (i.e. including the external
current ${\bf J}$). The result depends on the currents $J$, and this will
permit us to perform functional differentiations in order to calculate the
correlation, as shown in Eq.(\ref{deffifi}). As a first step in obtaining
the explicit form of ${\it Z}_J$, the perturbed soliton solutions depending
on ${\bf J}$ must be introduced in the expression of the action $S_J$ .
After that we perform the expansion of the functions $\varphi $ and $\chi $
around the coherent solution, 
\begin{eqnarray}
\varphi &=&\varphi _{Js}+\delta \varphi  \label{exp1} \\
\chi &=&\chi _{Js}+\delta \chi  \nonumber
\end{eqnarray}
This gives 
\begin{eqnarray*}
{\it Z}_J &=&\exp \left( iS_{Js}\right) \int D\left[ \delta \varphi \right] D%
\left[ \delta \chi \right] \\
&&\;\;\;\;\;\;\;\times \exp \left\{ \int d{\bf x}^{\prime }dt^{\prime
}\;\delta \chi \left( {\bf x}^{\prime },t^{\prime }\right) \,\left( \left. 
\frac{\delta ^2\widehat{{\it O}}}{\delta \varphi \delta \chi }\right|
_{\varphi _{Js},\chi _{Js}}\right) {\it \,}\delta \varphi \left( {\bf x}%
^{\prime },t^{\prime }\right) \right\}
\end{eqnarray*}
or 
\begin{equation}
{\it Z}_J=\exp \left( iS_{Js}\right) \frac 1{2^ni^n}\left( 2\pi \right)
^{n/2}\left( \det \left. \frac{\delta ^2\widehat{{\it O}}}{\delta \varphi
\delta \chi }\right| _{\varphi _{Js},\chi _{Js}}\right) ^{-1/2}
\label{exp22}
\end{equation}
since the integral is gaussian \cite{Amit}. The determinant is calculated
using the eigenvalues 
\begin{equation}
\left( \left. \frac{\delta ^2\widehat{{\it O}}}{\delta \varphi \delta \chi }%
\right| _{\varphi _{Js},\chi _{Js}}\right) \psi _n\left( {\bf x},t\right)
=\lambda _n\psi _n\left( {\bf x},t\right)  \label{exp3}
\end{equation}
and 
\begin{equation}
\det \left( \left. \frac{\delta ^2\widehat{{\it O}}}{\delta \varphi \delta
\chi }\right| _{\varphi _{Js},\chi _{Js}}\right) =\prod_n\lambda _n
\label{exp4}
\end{equation}

Since the action is invariant to the arbitrary position of the centre of the
soliton there are directions in the function space where the fluctuations
are not bounded and in particular are not Gaussian. This requires the
introduction of a set of collective coordinates and after a change of
variables the functional integrations along those particular directions are
replaced by usual integrations over the colective variables, with inclusion
of Jacobian factors. The zero eigenvalues of the determinant (corresponding
to the {\it zero modes}) are excluded in this way. We shall avoid this
complicated problem and assume a given position for the centre of the vortex.

\section{Application to the vortex solution of the nonlinear drift wave}

\subsection{The action functional}

In order to adimensionalize the equation (\ref{eqh1}) we introduce the space
and time scales $t\rightarrow \Omega ^{-1}t$ and $y\rightarrow \rho _sy$ and
the equation becomes 
\begin{equation}
\left( 1-\nabla _{\perp }^2\right) \frac{\partial \varphi }{\partial t}%
+\left( \frac{v_d}{\Omega \rho _s}\right) \frac{\partial \varphi }{\partial y%
}-\left( \frac{v_d}{\Omega \rho _s}\right) \varphi \frac{\partial \varphi }{%
\partial y}=0  \label{equa3}
\end{equation}
For simplicity of notation we keep the symbol $v_d$ for the adimensional
velocity $\left( \frac{v_d}{\Omega \rho _s}\right) $. The equation does not
change of form but now all variables are adequately normalized and the {\it %
action} 
\begin{equation}
S=\int dydt\,{\it L}_{\varphi}  \label{equa4}
\end{equation}
is also adimensional.

We have to calculate explicitely the scalar function $\chi $. Based on the
extended knowledge developed in field theory it seems reasonable to assume
that this function represents the generalization of the functions which have
the opposite evolution compared to $\varphi $: if $\varphi $ evolves toward
infinite time, then $\chi $ comes from infinite time toward the initial
time. If $\varphi $ diffuses then $\chi $ anti-diffuses (see Ref.\cite
{florinmadi3}). The general characteristics of this behaviour suggest to
represent $\chi $ as the object with the opposite topology than $\varphi $.
If $\varphi $ has a certain topological class, then $\chi $ has the opposite
topological class. If $\varphi $ is an instanton then $\chi $ is an
anti-instanton. In our case: if $\varphi $ is the vortex solution, then $%
\chi $ must the ``anti-vortex'' solution, with everywhere opposite vorticity
compared to $\varphi $. In our case of a single vortex, $\chi $ must simply
be a negative vortex.

In general terms, the direct ({\it i.e. }the vortex + random drift waves)
solution $\varphi $ arises from an initial perturbation which evolving in
time breaks into several distinct vortices (solitons) and a tail of drift
waves, as shown by the Inverse Scattering Method. The functionally
conjugated (``regressive'') function $\chi $ is at $t=\infty $ a collection
of vortices and drift wave turbulence which evolving backward in time,
toward $t=0$, coalesce and build up into a single perturbation, the same as
the initial condition of $\varphi $. We can restrict our analysis to the
time range where the two functions has similar patterns (but opposite) which
simply means to chose the time interval far from the initial and asymptotic
limits. As shown by analytical and numerical studies, the vortices (positive
and negative) are robust patterns and the time evolution simply consists of
translations without decay.{\bf \ }In conclusion we can take for the time
range far from the boundaries $t=0$ and $t=T$: 
\begin{equation}
\chi =-\varphi  \label{aps1}
\end{equation}

To see this more clearly, we write down the action and then the
Euler-Lagrange equations, with the current ${\bf J}$ included. 
\begin{equation}
S_{J}\left[ \chi ,\varphi \right] =\int_{0}^{L}dy\int_{0}^{T}dt\,{\it L}%
_{J,\varphi }  \label{aps2}
\end{equation}
with the notation 
\begin{equation}
{\it L}_{J,\varphi }=\chi \left[ \left( 1-\nabla _{\perp }^{2}\right) \frac{%
\partial \varphi }{\partial t}+v_{d}\frac{\partial \varphi }{\partial y}%
-v_{d}\varphi \frac{\partial \varphi }{\partial y}\right] +J_{\varphi
}\varphi +J_{\chi }\chi  \label{aps3}
\end{equation}
When performing integrations by parts the boundary conditions of the two
functions prevents us from taking the integrals of exact differentials as
vanishing, but this just produces terms which do not contribute to the
determination of the solution of extremum. We shall first change the Eq.(\ref
{aps3}) such as to obtain by functional extremization an (Euler-Lagrange)
equation for the function $\chi $: 
\begin{equation}
{\it L}_{J,\varphi }^{(1)}=\chi \frac{\partial \varphi }{\partial t}-\left[
\nabla _{\perp }^{2}\chi \right] \frac{\partial \varphi }{\partial t}%
+v_{d}\chi \frac{\partial \varphi }{\partial y}-v_{d}\chi \varphi \frac{%
\partial \varphi }{\partial y}+J_{\varphi }\varphi +J_{\chi }\chi
\label{aps7}
\end{equation}
Now we write the condition of extremum for the action functional and obtain
the Euler-Lagrange equation 
\begin{equation}
\frac{d}{dt}\frac{\delta {\it L}_{J,\varphi }^{(1)}}{\delta \left( \frac{%
\partial \varphi }{\partial t}\right) }+\frac{d}{dx}\frac{\delta {\it L}%
_{J,\varphi }^{(1)}}{\delta \left( \frac{\partial \varphi }{\partial x}%
\right) }+\frac{d}{dy}\frac{\delta {\it L}_{J,\varphi }^{(1)}}{\delta \left( 
\frac{\partial \varphi }{\partial y}\right) }-\frac{\delta {\it L}%
_{J,\varphi }^{(1)}}{\delta \varphi }=0  \label{aps9}
\end{equation}
This equation can be written 
\begin{equation}
\left( 1-\nabla _{\perp }^{2}\right) \frac{\partial \chi }{\partial t}+v_{d}%
\frac{\partial \chi }{\partial y}-v_{d}\varphi \frac{\partial \chi }{%
\partial y}=J_{\varphi }  \label{aps11}
\end{equation}
An equivalent form of the action is 
\begin{equation}
S_{J}\left[ \chi ,\varphi \right] =\int_{0}^{L}dy\int_{0}^{T}dt\,{\it L}%
_{J,\chi }  \label{aps13}
\end{equation}
with 
\begin{equation}
{\it L}_{J,\chi }=-\varphi \frac{\partial \chi }{\partial t}+\left( \nabla
_{\perp }^{2}\varphi \right) \frac{\partial \chi }{\partial t}-v_{d}\varphi 
\frac{\partial \chi }{\partial y}+v_{d}\frac{\varphi ^{2}}{2}\frac{\partial
\chi }{\partial y}+J_{\varphi }\varphi +J_{\chi }\chi  \label{aps14}
\end{equation}
The equation Euler-Lagrange for the function $\chi $ is obtained from the
extremum condition on the functional Eq.(\ref{aps13}) 
\begin{equation}
\frac{d}{dt}\frac{\delta {\it L}_{J,\chi }}{\delta \left( \frac{\partial
\chi }{\partial t}\right) }+\frac{d}{dx}\frac{\delta {\it L}_{J,\chi }}{%
\delta \left( \frac{\partial \chi }{\partial x}\right) }+\frac{d}{dy}\frac{%
\delta {\it L}_{J,\chi }}{\delta \left( \frac{\partial \chi }{\partial y}%
\right) }-\frac{\delta {\it L}_{J,\chi }}{\delta \chi }=0  \label{aps20}
\end{equation}
This equation reproduces the nonlinear vortex equation with an inhomogeneous
term: 
\begin{equation}
\left( 1-\nabla _{\perp }^{2}\right) \frac{\partial \varphi }{\partial t}%
+v_{d}\frac{\partial \varphi }{\partial y}-v_{d}\varphi \frac{\partial
\varphi }{\partial y}=-J_{\chi }  \label{aps21}
\end{equation}
Comparing the {\it homogeneous }equations (\ref{aps11}) (with $J_{\varphi
}=0 $) and (\ref{aps21}) (with $J_{\chi }=0$) we see that 
\begin{equation}
\chi =-\varphi  \label{aps22}
\end{equation}
is indeed the solution of the {\it homogeneous }equation (\ref{aps11}) {\it %
i.e. }the{\it \ }negative vortex is the solution for $\chi $.

We must remember that the ``external'' currents are arbitrary and later,
after functional differentation, they will be taken zero. This allows us to
start from the configurations given by the {\it homogeneous} equations and
Eq.(\ref{aps22}) and study the small changes using perturbative methods
developed in the framework of the Inverse Scattering Transform. We will only
use the current $J_{{\varphi}} $ which will be denoted $J$ and already take $%
J_{\chi} =0$.

The final form of the action which will be used later in this work is 
\begin{equation}
S_J\left[ \chi ,\varphi \right] =\int_0^Ldy\int_0^Tdt\,\left\{ \chi \left(
1-\nabla _{\perp }^2\right) \frac{\partial \varphi }{\partial t}+v_d\chi 
\frac{\partial \varphi }{\partial y}-v_d\chi \varphi \frac{\partial \varphi 
}{\partial y}+J\varphi \right\}  \label{aps23}
\end{equation}

\subsection{The condition of extremum of the action functional}

The Euler - Lagrange equations for the two functions $\chi $ and $\varphi $
are obtained from the first functional derivative of the action $S_J$: $%
\delta S_J/\delta \chi =0$ and $\delta S_J/\delta \varphi =0$. The first
equation (which is the original equation) has the solution (\ref{eqh7}). It
does not depend on the current $J$ (since the corresponding current $J_\chi $
has been taken zero). However, for uniformity of notation we shall write $%
\varphi _{Js}$%
\begin{equation}
\varphi _{Js}(x,y,t)\equiv \varphi _s\left( x,y,t\right)  \label{fijs}
\end{equation}

The second Euler-Lagrange equation is the equation for $\chi $, with the
inhomogeneous term given by the current $J$: 
\begin{equation}
\left( 1-\nabla _{\perp }^2\right) \frac{\partial \chi }{\partial t}+v_d%
\frac{\partial \chi }{\partial y}-v_d\varphi \frac{\partial \chi }{\partial y%
}=J  \label{eqhij}
\end{equation}
The solution is: 
\begin{equation}
\chi _{Js}\left( x,y,t\right) =-\,\varphi _s\left( x,y,t\right) +\widetilde{%
\chi }_J\left( x,y,t\right)  \label{hijs}
\end{equation}
where $-\,\varphi _s\left( x,y,t\right) $ represents the ``free'' solution
of the variational equation, {\it i.e.} the negative vortex (anti-soliton)
and $\widetilde{\chi }_J\left( x,y,t\right) $ is the small modification
induced by an inhomogeneous small term, $J\left( x,y,t\right) $. Since the
function $\widetilde{\chi }_J\left( x,y,t\right) $ is the perturbation of
the negative-vortex solution we will use the equation (\ref{aps21}) but with
the opposite current ({\it i.e.} $-J$ instead of $J$), as (\ref{eqhij})
requires.

\subsection{Second order functional expansion and the eigenvalue problem for
the calculation of the Determinant}

Now we shall expand the action $S_J\left[ \varphi \right] $ to second order
around the saddle-point solution. Write 
\begin{eqnarray}
\varphi &=&\varphi _{Js}+\delta \varphi  \label{expan2} \\
\chi &=&\chi _{Js}+\delta \chi  \nonumber
\end{eqnarray}
where the function $\left( \delta \varphi ,\delta \chi \right) $ is a small
difference from the extremum solution.The expanded form of the action will
be written: 
\[
S_J\left[ \chi ,\varphi \right] =S_J\left[ \varphi _{Js},\chi _{Js}\right]
+\frac 12\left( \left. \frac{\delta ^2S_J}{\delta \varphi \delta \chi }%
\right| _{\varphi _{Js},\chi _{Js}}\right) \delta \varphi \delta \chi 
\]
where obviously the absence of the linear term is due to the fact that $%
\left( \varphi _{Js},\chi _{Js}\right) $ is the solution at the extremum and 
\begin{eqnarray}
S_J\left[ \varphi _{Js},\chi _{Js}\right] &=&\int_0^Ldy\int_0^Tdt\left[ \chi
_{Js}\frac{\partial \varphi _{Js}}{\partial t}-\left( \nabla _{\perp }^2\chi
_{Js}\right) \frac{\partial \varphi _{Js}}{\partial t}\right.  \label{expan4}
\\
&&+v_d\chi _{Js}\frac{\partial \varphi _{Js}}{\partial y}-v_d\chi
_{Js}\varphi _{Js}\frac{\partial \varphi _{Js}}{\partial y}+J\varphi _{Js} 
\nonumber
\end{eqnarray}
Few manipulations are necessary to make the second functional variation of $%
S_J$ symmetric in $\delta \varphi $ and $\delta \chi $. Again this will
imply boundary terms, but these are now zero since the variations $\delta
\varphi $ and $\delta \chi $ vanishes at the limits of the space-time
domain, by definition. The transformations are simply integrations by parts
and give 
\begin{equation}
\frac 12\,\delta \chi \,\left( \left. \frac{\delta ^2S_J}{\delta \varphi
\delta \chi }\right| _{\varphi _{Js},\chi _{Js}}\right) \delta \varphi
=\frac 12\mathrel{\mathop{\stackrel{\frown }{%
\begin{array}{ll}
\delta \varphi & \delta \chi
\end{array}
}}\limits_{\smile }}%
\left( 
\begin{array}{cc}
\widehat{\gamma } & -\widehat{\alpha }-\widehat{\beta } \\ 
\widehat{\alpha }-\widehat{\beta } & 0
\end{array}
\right) \left( 
\begin{array}{l}
\delta \varphi \\ 
\delta \chi
\end{array}
\right)  \label{oper}
\end{equation}
where 
\begin{eqnarray}
\widehat{\alpha } &=&\left( 1-\nabla _{\perp }^2\right) \frac \partial
{\partial t}+\,v_d\frac \partial {\partial y}-v_d\left( \frac{\partial
\varphi _{Js}}{\partial y}\right)  \label{oper2} \\
\widehat{\beta } &=&\frac 12v_d\left( \frac{\partial \varphi _{Js}}{\partial
y}\right)  \nonumber \\
\widehat{\gamma } &=&-2v_d\chi _{Js}\frac \partial {\partial y}  \nonumber
\end{eqnarray}
In the generating functional of the correlations, the expansion gives, after
performing the Gaussian integral: 
\begin{equation}
{\it Z}_J=\exp \left( iS_J\right) \frac 1{i^n}\left( 2\pi \right) ^{n/2}%
\left[ \det \left( \left. \frac{\delta ^2S_J}{\delta \varphi \delta \chi }%
\right| _{\varphi _{Js},\chi _{Js}}\right) \right] ^{-1/2}  \label{oper3}
\end{equation}
As stated before, the $\det $ will be calculated as the product of the
eigenvalues $\lambda _n$%
\begin{equation}
\det \left( \left. \frac{\delta ^2S_J}{\delta \varphi \delta \chi }\right|
_{\varphi _{Js},\chi _{Js}}\right) =\prod_n\lambda _n  \label{oper4}
\end{equation}
We must find the eigenvalues of the differential operator appearing in Eq.(%
\ref{oper}): 
\begin{equation}
\left( 
\begin{array}{cc}
\widehat{\gamma } & -\widehat{\alpha }-\widehat{\beta } \\ 
\widehat{\alpha }-\widehat{\beta } & 0
\end{array}
\right) \left( 
\begin{array}{c}
\psi _n^\varphi \\ 
\psi _n^\chi
\end{array}
\right) =\lambda _n\left( 
\begin{array}{c}
\psi _n^\varphi \\ 
\psi _n^\chi
\end{array}
\right)  \label{oper5}
\end{equation}
which gives the following equation 
\begin{equation}
\left[ \widehat{\gamma }-\frac 1{\lambda _n}\left( \widehat{\alpha }^2+%
\widehat{\beta }\widehat{\alpha }-\widehat{\alpha }\widehat{\beta }-\widehat{%
\beta }^2\right) \right] \psi _n^\varphi =\lambda _n\psi _n^\varphi
\label{oper6}
\end{equation}
The functions $\delta \varphi \left( y,t\right) $ and $\delta \chi \left(
y,t\right) $ represent the differences between the solutions at extremum
(solitons) and other functions which are in a neighbourhood (in the function
space) of the solitons. According to the discussion above, the functions
which are ``close'' to the solitons, for the Flierl-Petviashvilli equation
are drift waves. For this reason the operator which represents the
dispersion ({\it i.e. }$\nabla _{\perp }^2$) will be replaced with its
simplest form , $-\overline{k}_{\perp }^2$ for these waves, with $\overline{k%
}_{\perp }$ representing an average normalized wavenumber for the pure drift
turbulence. However, the operator will be retained when applied on the
functions related to solitons, since these solutions owe their existence to
the balance of nonlinearity and dispersion.The following detailed
expressions are obtained for the operators involved in this equation: 
\begin{eqnarray}
\widehat{\alpha }^2 &=&\left[ \left( 1-\nabla _{\perp }^2\right) \frac
\partial {\partial t}+\,v_d\frac \partial {\partial y}-v_d\left( \frac{%
\partial \varphi _{Js}}{\partial y}\right) \right] ^2  \label{oper7} \\
&=&\left( 1+\overline{k}_{\perp }^2\right) ^2\left( \frac \partial {\partial
t}\right) ^2  \nonumber \\
&&-v_d\left[ \left( 1-\nabla _{\perp }^2\right) \frac{\partial \varphi _{Js}%
}{\partial t}\right] \frac \partial {\partial y}  \nonumber \\
&&+v_d\left( 1-\varphi _{Js}\right) \left( 1+\overline{k}_{\perp }^2\right) 
\frac{\partial ^2}{\partial t\partial y}+v_d\left( 1-\varphi _{Js}\right)
\left( 1+\overline{k}_{\perp }^2\right) \frac{\partial ^2}{\partial
y\partial t}  \nonumber \\
&&-v_d^2\left( 1-\varphi _{Js}\right) \left( \frac{\partial \varphi _{Js}}{%
\partial y}\right) \frac \partial {\partial y}  \nonumber \\
&&+v_d^2\left( 1-\varphi _{Js}\right) ^2\frac{\partial ^2}{\partial y^2} 
\nonumber
\end{eqnarray}
\begin{equation}
\widehat{\beta }\widehat{\alpha }=\frac 12v_d\left( \frac{\partial \varphi
_{Js}}{\partial y}\right) \left( 1+\overline{k}_{\perp }^2\right) \frac
\partial {\partial t}+\frac 12v_d^2\left( \frac{\partial \varphi _{Js}}{%
\partial y}\right) \left( 1-\varphi _{Js}\right) \frac \partial {\partial y}
\label{oper8}
\end{equation}
\begin{eqnarray}
\widehat{\alpha }\widehat{\beta } &=&\frac 12v_d\left[ \left( 1-\nabla
_{\perp }^2\right) \frac{\partial ^2\varphi _{Js}}{\partial t\partial y}%
\right]  \label{oper9} \\
&&+\frac 12v_d^2\left( 1-\varphi _{Js}\right) \left( \frac{\partial
^2\varphi _{Js}}{\partial y^2}\right) +\frac 12v_d^2\left( 1-\varphi
_{Js}\right) \left( \frac{\partial \varphi _{Js}}{\partial y}\right) \frac
\partial {\partial y}  \nonumber \\
&&+\frac 12v_d\left( \frac{\partial \varphi _{Js}}{\partial y}\right) \left(
1+\overline{k}_{\perp }^2\right) \frac \partial {\partial t}  \nonumber
\end{eqnarray}
\begin{equation}
\widehat{\beta }^2=\frac 14v_d^2\left( \frac{\partial \varphi _{Js}}{%
\partial y}\right) ^2  \label{oper10}
\end{equation}
The square brakets are used to underline that the differential operators are
not acting outside and the only operation is multiplication. We use the
equation verified by $\varphi _{Js}$ to make the following replacement 
\begin{equation}
\left[ \left( 1-\nabla _{\perp }^2\right) \frac{\partial \varphi _{Js}}{%
\partial t}\right] =-v_d\left( 1-\varphi _{Js}\right) \left( \frac{\partial
\varphi _{Js}}{\partial y}\right)  \label{oper11}
\end{equation}
The equation becomes 
\begin{eqnarray}
&&-\left\{ \left( 1+\overline{k}_{\perp }^2\right) ^2\frac{\partial ^2}{%
\partial t^2}+2\left( 1+\overline{k}_{\perp }^2\right) v_d\left( 1-\varphi
_{Js}\right) \frac{\partial ^2}{\partial y\partial t}\right.  \label{oper12}
\\
&&\left. \,\,\,\;\;\;\;\;\;+v_d^2\left( 1-\varphi _{Js}\right) ^2\frac{%
\partial ^2}{\partial y^2}-\frac 34v_d^2\left( \frac{\partial \varphi _{Js}}{%
\partial y}\right) ^2\right\} \psi _n^\varphi  \nonumber \\
&&+\lambda _n\left( -2v_d\chi _{Js}\frac \partial {\partial y}\right) \psi
_n^\varphi  \nonumber \\
&=&\lambda _n^2\;\psi _n^\varphi  \nonumber
\end{eqnarray}

We now take into account the propagating nature of the drift waves and make
the change of variables $t\rightarrow t$ and $y\rightarrow y-v_dt$ {\it i.e.}
we change to the system of reference moving with the diamagnetic velocity.
We simplify the equation assuming that the most important space-time
variation is wave-like and replace $\frac \partial {\partial t}=-v_d\frac
\partial {\partial y}$. By this change of variables the soliton will not be
at rest in the new reference system, but it will move very slowly since we
have assumed that $u\gtrsim v_d$. We make another approximation by
neglecting the slow motion of the soliton. This restricts us to the
wavenumber spectrum but considerably simplifies the calculations. The space
variable which will be denoted again $y$ measures the space from the fixed
center of the soliton, in the moving system. The difference between the KdV
soliton, which is one-dimensional and depends exclusively on $y$ and the
vortex which is a two-dimensional structure will be considered in the
simplest form as described by the estimation of Meiss and Horton for the $x$
- extension of the vortex. For convenience we suppress the index $n$ and
replace $\psi _n^\varphi $ by $q$. 
\begin{eqnarray}
&&\left\{ \left[ \left( 1+\overline{k}_{\perp }^2\right) v_d-v_d\left(
1-\varphi _{Js}\right) \right] ^2\frac{\partial ^2}{\partial y^2}\right.
\label{lvp} \\
&&\;\;\;+\left( 2\lambda \,v_d\chi _{Js}\right) \frac \partial {\partial y} 
\nonumber \\
&&\;\;\;\left. +\left[ \lambda ^2-\frac 34v_d^2\left( \frac{\partial \varphi
_{Js}}{\partial y}\right) ^2\right] \right\} \;q  \nonumber \\
&=&0  \nonumber
\end{eqnarray}

We have a suggestive confirmation that the generating function ${\it Z}_J$
(via the action $S_J$) potentially contains configurations of the system
consisting of simple drift waves. A perturbation consisting of drift waves
and propagating with the diamagnetic velocity $v_d$ is an approximate
solution of the original equation for small amplitude ({\it i.e.} small
nonlinearity). Due to its particular structure, the Martin-Siggia-Rose
action functional is exactly zero when calculated with the exact solution,
in the absence of any external current $J$. The action expanded to the
second order then gives, for no vortex ($\varphi _{Js}=0$, $\chi _{Js}=0$) 
\begin{equation}
\left[ \frac{\partial ^2}{\partial y^2}+\left( \frac \lambda {\overline{k}%
_{\perp }^2v_d}\right) ^2\right] \,q=0  \label{oper14}
\end{equation}
which implies periodic oscillations in the space variable $y$ with (recall
that everything is adimensional) 
\begin{equation}
\lambda =k_y\,v_d\left( \overline{k}_{\perp }^2\right)  \label{oper15}
\end{equation}

Returning to the equation (\ref{lvp}) we write it in the following form 
\begin{equation}
\left( \frac{\partial ^2}{\partial y^2}+A\frac \partial {\partial
y}+B\right) q=0  \label{oper16}
\end{equation}
where 
\begin{eqnarray}
A &\equiv &\frac{2\lambda }{v_d}\frac{\chi _{Js}}{\left( \overline{k}_{\perp
}^2+\varphi _{Js}\right) ^2}  \label{oper17} \\
B &\equiv &\frac{\frac{\lambda ^2}{v_d^2}-\frac 34\left( \frac{\partial
\varphi _{Js}}{\partial y}\right) ^2}{\left( \overline{k}_{\perp }^2+\varphi
_{Js}\right) ^2}  \nonumber
\end{eqnarray}
Now we make the standard transformation of the unknown function 
\begin{equation}
q=w\exp \left( -\frac 12\int^yA\left( y^{\prime }\right) dy^{\prime }\right)
\label{oper18}
\end{equation}
and obtain 
\begin{equation}
w^{\prime \prime }+\left( B-\frac{A^{\prime }}2-\frac{A^2}4\right) w=0
\label{oper19}
\end{equation}
where prime means derivation with respect to $y$. After replacing the two
extremum solutions $\varphi _{Js}$ and $\chi _{Js}$ from equations (\ref
{fijs}) and (\ref{hijs}) this equation is written in the following form, to
exhibit the dependence on $\lambda $: 
\begin{equation}
w^{\prime \prime }+\left( \lambda ^2\,t_1+\lambda \,t_2+t_3\right) w=0
\label{wsec}
\end{equation}
with the notations 
\begin{equation}
t_1(y)\equiv \frac 1{v_d^2}\frac{h^2-\varphi _s^2}{h^4}+\frac 2{v_d}\frac{%
\varphi _s}{h^4}\;\widetilde{\chi }_J  \label{oper20}
\end{equation}
\begin{equation}
t_2\left( y\right) \equiv -\frac 1{v_d}\left( \frac{\partial \varphi _s}{%
\partial y}\right) \frac{2c-h}{h^3}+\frac 2{v_d}\frac{\left( \frac{\partial
\varphi _s}{\partial y}\right) }{h^3}\,\widetilde{\chi }_J-\frac 1{v_d}\frac
1{h^2}\left( \frac{\partial \widetilde{\chi }_J}{\partial y}\right)
\label{oper21}
\end{equation}
\begin{equation}
t_3\left( y\right) \equiv -\frac 34\frac 1{h^2}\left( \frac{\partial \varphi
_s}{\partial y}\right) ^2  \label{oper22}
\end{equation}
and 
\begin{eqnarray}
c &\equiv &\overline{k}_{\perp }^2  \label{oper23} \\
h &=&c+\varphi _s  \nonumber
\end{eqnarray}

The functions $t_{i}\left( y\right) $ are represented for $i=1,2$ in Figure
2 and 3. The function 
\begin{equation}
U\left( \lambda ;y\right) \equiv \lambda ^{2}\,t_{1}+\lambda \,t_{2}+t_{3}
\label{oper24}
\end{equation}
has singularities at the points where $h$ vanishes. We introduce the
notation $y_{h}$ for the location of the singularities, taking into account
the symmetry around $y=0$, the centre of the soliton 
\begin{equation}
h\left( \pm y_{h}\right) =0.  \label{oper25}
\end{equation}
Since the soliton is very localized, the function $U$ has very fast
variations close to the singularities. The slow variation of the function $%
U\left( \lambda ;y\right) $ over most of the space interval $\left(
-L/2,+L/2\right) $ becomes very fast due to the growth of the absolute
values of $t_{1}$, $t_{2}$ and $t_{3}$ near $\pm y_{h}$, on spatial
intervals having an extension of the order of the spatial unit, {\it i.e.} $%
\rho _{s}$ in physical terms. Since the physical model leading to our
original equation cannot accurately describe the physical processes at such
scales, we shall adopt the simplest approximation of $U$, assuming that it
reaches infinite absolute value at points which are located whithin a
distance of $\rho _{s}$ of the actual positions of the singularities, $\pm
y_{h}$. We have checked that the exact position of the assumed {\it infinite}
value of $U$ has no significant impact on the final results, which can be
explained by observing that $t_{1,2,3}$ will be integrated on. The total
space interval is now divided into three domains: $\left( -L/2,-y_{h}\right) 
$ (external left), $\left( -y_{h},y_{h}\right) $ (internal) and $\left(
y_{h},L/2\right) $ (external right). Here ``internal'' and ``external''
refer to the region approximatly occupied by the soliton. The form of the
function $U$ imposes the function $w$ to vanish at the limits of these
domains. In a more general perspective, the fact that $w$ behaves
independently on each domain has a consequence with statistical mechanics
interpretation: the generating functional (similar to any partition
function) is obtained by integrating over the full space of the system's
physical configurations and behaves multiplicatively for any splitting of
the whole function space into disjoint subspaces. In particular the
functional integration over the space of functions $\delta \varphi $ and $%
\delta \chi $ actually consists of three functional integrations over the
disjoint function subspaces corresponding to the three spatial domains. The
fact that our physical model is restricted to spatial scales larger than $%
\rho _{s}$ necessarly has an impact on the maximum number of eigenvalues $%
\lambda _{n}$ that should be retained in the infinite product giving the
determinant, but we shall not need to use this limitation.

For absolute values of the parameter $\lambda $ greater than unity (which
will be confirmed {\it a posteriori}, by the expressions (\ref{oper28}) and (%
\ref{oper37}) below), the three terms in the expression of $U$ have very
different contributions. The terms $t_3$ is practically negligible, and the
term with $t_1$ is always much greater than $t_2$ in absolute value. In the
following we consider separately the three domains.

On the ``external left'' domain, the function $t_1$ is positive. If we fix
at zero the amplitude and the phase of $w$ at the limit $-L/2$ the condition
that the solution vanishes at $-y_h$ gives, for $\lambda $ real, 
\begin{equation}
\int_{-L/2}^{-y_h}dy^{\prime }\,\left( \lambda ^2\,t_1+\lambda
\,t_2+t_3\right) ^{1/2}=2\pi n  \label{oper26}
\end{equation}
In the integrand, the first term is factorized and, taking into account the
relative magnitude of the terms, we expand the square root and obtain 
\begin{equation}
\lambda _n^l\alpha _1+\beta _1+\frac{\gamma _1}{\lambda _n^l}=2\pi n
\label{oper27}
\end{equation}
{\it i.e.} 
\begin{equation}
\lambda _n^l=\frac{2\pi n}{\alpha _1}\left( 1-\frac{\beta _1/\left( 2\pi
\right) }n\right)  \label{oper28}
\end{equation}
where 
\begin{equation}
\alpha _1=\int_{-L/2}^{-y_h}dy^{\prime }\,\sqrt{t_1\left( y^{\prime }\right) 
}  \label{oper29}
\end{equation}
\begin{equation}
\beta _1=\int_{-L/2}^{-y_h}dy^{\prime }\frac{t_2\left( y^{\prime }\right) }{%
\sqrt{t_1\left( y^{\prime }\right) }}  \label{oper30}
\end{equation}
\begin{equation}
\gamma _1=\int_{-L/2}^{-y_h}dy^{\prime }\frac{t_3\left( y^{\prime }\right) }{%
\sqrt{t_1\left( y^{\prime }\right) }}  \label{oper31}
\end{equation}
and $\gamma _1$ has been neglected. We note that $\beta _1$ is positive.

On the ``external right'' domain the function $t_1$ is positive but $t_2$ is
negative. The condition on the phase is 
\begin{equation}
\int_{y_h}^{L/2}dy^{\prime }\left( \lambda ^2\,t_1+\lambda \,t_2+t_3\right)
^{1/2}=2\pi n^{\prime }  \label{oper32}
\end{equation}
and introduce similar notations 
\begin{equation}
\alpha _2=\int_{y_h}^{L/2}dy^{\prime }\sqrt{t_1\left( y^{\prime }\right) }%
=\alpha _1  \label{oper33}
\end{equation}
\begin{equation}
\beta _2=\int_{y_h}^{L/2}dy^{\prime }\frac{t_2\left( y^{\prime }\right) }{2%
\sqrt{t_1\left( y^{\prime }\right) }}=-\beta _1  \label{oper34}
\end{equation}
\begin{equation}
\gamma _2=\int_{y_h}^{L/2}dy^{\prime }\frac{t_3\left( y^{\prime }\right) }{2%
\sqrt{t_1\left( y^{\prime }\right) }}  \label{oper35}
\end{equation}
The equation then becomes 
\begin{equation}
\lambda _{n^{\prime }}^r\,\alpha _2+\beta _2+\frac{\gamma _2}{\lambda
_{n^{\prime }}^r}=2\pi n^{\prime }  \label{oper36}
\end{equation}
or 
\begin{equation}
\lambda _{n^{\prime }}^r=\frac{2\pi n^{\prime }}{\alpha _2}\left( 1+\frac{%
\beta _1/\left( 2\pi \right) }{n^{\prime }}\right)  \label{oper37}
\end{equation}

The infinite product of eigenvalues gives, for the ``external'' region \cite
{grrj-8.322}: 
\begin{eqnarray}
\prod_n\lambda _n^l\prod_{n^{\prime }}\lambda _{n^{\prime }}^r
&=&\prod_n\left( \frac{2\pi n}{\alpha _1}\right) ^2\,\prod_n\left( 1-\frac{%
\beta _1^2/\left( 2\pi \right) ^2}{n^2}\right)  \label{oper38} \\
&=&\frac{\sin \left( \beta _1/2\right) }{\beta _1/2}\;\prod_n\left( \frac{%
2\pi n}{\alpha _1}\right) ^2\,  \nonumber
\end{eqnarray}

In the ``internal'' region, the function $t_1$ is negative. The relations
between the magnitudes of the absolute values of the functions $t_1$, $t_2$
and $t_3$ are preserved. Then $\lambda $ will be complex. Due to the
anti-symmetry of the function $t_2$ we can suppose that the unknown function 
$w$ takes zero value at $y=0$. We introduce the notations 
\begin{equation}
\alpha _c=\int_0^{y_h}dy^{\prime }\sqrt{-t_1\left( y^{\prime }\right) }
\label{oper39}
\end{equation}
\begin{equation}
\beta _c=\int_0^{y_h}dy^{\prime }\frac{t_2\left( y^{\prime }\right) }{2\sqrt{%
-t_1\left( y^{\prime }\right) }}  \label{oper40}
\end{equation}
\begin{equation}
\gamma _c=\int_0^{y_h}dy^{\prime }\frac{t_3\left( y^{\prime }\right) }{2%
\sqrt{-t_1\left( y^{\prime }\right) }}  \label{oper41}
\end{equation}
which are {\it real} numbers. The condition 
\begin{equation}
\lambda _n^i\alpha _c+\beta _c+\frac{\gamma _c}{\lambda _n^i}=2\pi in
\label{oper42}
\end{equation}
gives (after neglecting $\gamma _c$) for the complex number $\lambda _n^i$: 
\begin{equation}
\lambda _n^i=\alpha _c^{-1}\left( 2\pi n\right) \left( 1+\frac{\beta _c^2}{%
\left( 2\pi \right) ^2n^2}\right) ^{1/2}\exp \left[ -i\,\arctan \left( \frac{%
2\pi n}{\beta _c}\right) \right]  \label{oper43}
\end{equation}
The infinite product of these eigenvalues is 
\begin{equation}
\prod_n\lambda _n^i=\prod_n\alpha _c^{-1}\left( 2\pi n\right) \exp \left[
-i\,\arctan \left( \frac{2\pi n}{\beta _c}\right) \right] \;\prod_n\left( 1+%
\frac{\beta _c^2/\left( 2\pi \right) ^2}{n^2}\right) ^{1/2}  \label{oper44}
\end{equation}
The number $\beta _c$ is smaller than unity and for large $n$ the argument
of the exponential will be more and more close to $-i\,\pi /2$. We make the
approximation that the exponential can be replaced with $-i$. Then we obtain 
\begin{equation}
\prod_n\lambda _n^i=\left[ \frac{\sinh \left( \beta _c/2\right) }{\beta _c/2}%
\right] ^{1/2}\prod_n\frac{\left( -i\right) 2\pi n}{\alpha _c}\,
\label{oper45}
\end{equation}

On the ``external'' regions the functions $t_1$, $t_2$ are not symmetrical
around the centre $y=0$ since the perturbed soliton develops a ``tail''
which is not symmetrical. However we take this perturbation to be small and
assume the same absolute value for the function $\beta _1$ on both external
domains.

We remark that we remain with two quantities in which all the functional
depencence on the current $J$ is packed: for ``exterior'' $\beta _1$
(hereafter denoted $\sigma $) and for ``interior'' $\beta _c$ (hereafter
denoted $\beta $). 
\begin{eqnarray}
{\it Z}_J &=&\exp \left( iS_J\right) \left( \prod_n\frac{\left( 2\pi \right) 
}i\right) \left[ \det \left( \left. \frac{\delta ^2S_J}{\delta \varphi
\delta \chi }\right| _{\varphi _{Js},\chi _{Js}}\right) \right] ^{-1/2}
\label{oper56} \\
&=&const\;\exp \left( iS_J\right) \,\left[ \frac{\beta /2}{\sinh \left(
\beta /2\right) }\right] ^{1/4}\left[ \frac{\sigma /2}{\sin \left( \sigma
/2\right) }\right] ^{1/2}  \nonumber
\end{eqnarray}
where 
\begin{equation}
const=\prod_n\left( \frac{\left( -i\right) \alpha _c}{2\pi n}\right) ^{1/2}%
\frac{\alpha _1}n  \label{oper57}
\end{equation}
will disappear after the normalizations required by the calculation of the
correlations (see below).

\section{Calculation of the correlations}

The two-point correlation can be obtained by a double functional
differentiation at the external current $J$. 
\[
\left\langle \varphi (y_{2})\varphi (y_{1})\right\rangle =\left. {\it Z}%
_{J}^{-1}\frac{\delta ^{2}{\it Z}_{J}}{i\delta J(y_{2})\,i\delta J\left(
y_{1}\right) }\right| _{J=0} 
\]
The main achivement of this approach is that it provides the explicit
expression of the generating functional. We introduce the notations 
\begin{equation}
A=A\left[ J\right] \equiv \left[ \frac{\beta /2}{\sinh \left( \beta
/2\right) }\right] ^{1/4}  \label{cor3}
\end{equation}
\begin{equation}
B=B\left[ J\right] \equiv \left[ \frac{\sigma /2}{\sin \left( \sigma
/2\right) }\right] ^{1/2}  \label{cor4}
\end{equation}
and drop the factor $const$; actually the latter depends on $\alpha _{1}$
and $\alpha _{c}$ and thus on the current $J$ and contributes to the
functional derivatives. However, taking a formal limit $N$ to the number of
factors in (\ref{oper57}) we find that the functional derivatives of $\alpha
_{1}$ and $\alpha _{c}$ give additive terms which vanish in the limit $%
N\rightarrow \infty $. Then we drop $const$ since it disappears after
dividing to ${\it Z}_{J}$ and taking $J\equiv 0$. In this way (\ref{oper56})
becomes 
\begin{equation}
{\it Z}_{J}=\exp \left( iS_{J}\right) \,A\,B  \label{cor5}
\end{equation}
We calculate the functional derivatives. 
\begin{equation}
\frac{\delta {\it Z}_{J}}{i\delta J\left( y_{1}\right) }=\left[ \frac{\delta
S_{J}}{\delta J\left( y_{1}\right) }+\frac{1}{A}\frac{\delta A}{i\delta
J\left( y_{1}\right) }+\frac{1}{B}\frac{\delta B}{i\delta J\left(
y_{1}\right) }\right] \exp \left( iS_{J}\right) \,A\,B  \label{cor6}
\end{equation}
We will also need the functional derivative at $J\left( y_{2}\right) $, with
a similar expression.The second derivative: 
\begin{eqnarray}
\left. {\it Z}_{J}^{-1}\frac{\delta ^{2}{\it Z}_{J}}{i\delta
J(y_{2})\,i\delta J\left( y_{1}\right) }\right| _{J=0} &=&\frac{\delta S_{J}%
}{\delta J\left( y_{2}\right) }\frac{\delta S_{J}}{\delta J\left(
y_{1}\right) }+\frac{\delta ^{2}S_{J}}{i\delta J\left( y_{2}\right) \delta
J\left( y_{1}\right) }  \label{cor7} \\
&&+\frac{1}{A}\frac{\delta A}{i\delta J\left( y_{2}\right) }\frac{\delta
S_{J}}{\delta J\left( y_{1}\right) }+\frac{1}{B}\frac{\delta B}{i\delta
J\left( y_{2}\right) }\frac{\delta S_{J}}{\delta J\left( y_{1}\right) } 
\nonumber \\
&&+\frac{1}{A}\frac{\delta A}{i\delta J\left( y_{1}\right) }\frac{\delta
S_{J}}{\delta J\left( y_{2}\right) }+\frac{1}{B}\frac{\delta B}{i\delta
J\left( y_{1}\right) }\frac{\delta S_{J}}{\delta J\left( y_{2}\right) } 
\nonumber \\
&&+\frac{1}{A}\frac{\delta A}{i\delta J\left( y_{1}\right) }\frac{1}{B}\frac{%
\delta B}{i\delta J\left( y_{2}\right) }+\frac{1}{A}\frac{\delta A}{i\delta
J\left( y_{2}\right) }\frac{1}{B}\frac{\delta B}{i\delta J\left(
y_{1}\right) }  \nonumber \\
&&+\frac{1}{A}\frac{\delta ^{2}A}{i\delta J\left( y_{2}\right) i\delta
J\left( y_{1}\right) }+\frac{1}{B}\frac{\delta ^{2}B}{i\delta J\left(
y_{2}\right) i\delta J\left( y_{1}\right) }  \nonumber
\end{eqnarray}
The detailed expressions of these terms are given in the Appendix. The terms
are calculated numerically using the detailed expressions of $\varphi _{s}$, 
$\frac{\partial \varphi _{s}}{\partial y}$, $\widetilde{\chi }_{J}$ and $%
\frac{\partial \widetilde{\chi }_{J}}{\partial y}$. 

The first term reproduces the self-correlation of the soliton and represents
the connection with the results of Ref.\cite{Pf25-82-1838}, with our
particular simplifications: single soliton and fixed (non-random) position
of its centre. As can easily be seen, the first order functional derivatives
of $S_{J}$ to the current $J$ reduce to the function $\varphi _{s}$
calculated in the corresponding points. The term with the double functional
derivative of the action represents the contribution to the self-correlation
of the soliton due to a statistical ensemble of initial conditions, without
drift waves. All mixed terms ({\it i.e.} containing both the action and one
of the factors $A$ or $B$) represent interaction between the perturbed
soliton and the drift waves. The terms containing exclusively the factors $A$
and/or $B$ refers to the drift waves {\it in the presence} of the perturbed
soliton.

\section{Discussion and conclusions}

The formulas obtained by functional differentiation of the generating
functional are complicated and a numerical calculation is necessary. We
chose a particular value of the soliton velocity (which also fixes its
amplitude): $u=1.725\,v_d$ and let the variables $y_1$ and $y_2$ sample the
one-dimensional volume of length $L=0.2\,m$. The physical parameters are
chosen such that $\rho _s\approx 10^{-3}m$ and $v_d\approx 571\;m/s$. We
recall that there are two particular symmetry limitations of our
calculation. (1) The soliton centre is assumed fixed (at $y=0$) , especially
for avoiding the complicated problem of the {\it zero modes}. (2) Due to the
asymmetry of the perturbed soliton tail the terms which results from the
functional differentiation are also asymmetric. These are only limitations
of our calculation and in no way reflect the reality of a isotropic motion
of many solitons in a real turbulent plasma. In order to see to what extent
our result can be useful for understatnding the (much more complicated) real
situation we will symmetrize these terms in the unique mode which is
accessible to our one-dimensional calculation, i.e. take into account the
mixing of perturbed solitons moving in the two directions on the line.

The amplitude of the modifications of the soliton depends on a parameter
which is the average time of interaction with the perturbation. This average
time is comparable with the time required to cross $L$ at a speed of $v_d$
and is limited since the growth of the perturbation cannot exceed the
soliton itself.

The figures {\it are conventional representations of functions of two
variables} $(y_{1},y_{2})$; they do not correspond to a two-dimensional
geometry. For this reason it is not expected to have {\it circular}{\bf \ }%
symmetry. The contributions to the correlation from the last two factors in
Eq.(\ref{cor7}) have amplitudes similar or less by a factor of few units,
compared to the pure soliton. The factors coming from ``internal'' part are
peaked and localized on the soliton extension while the ``external'' part
gives terms oscillating on $\left( y_{1},y_{2}\right) $. In wavenumber
space, there are contributions to both low-$k$ and high-$k$ regions. The
spectrum of an unperturbed soliton is smooth and monotonously decreasing
from the peak value at ${\bf k=0}$. Fig.5 shows much
more structure. In the low-$k$ part there are many local peaks, an effective
manifestation of the periodic character of the terms (as shown by (\ref
{oper38}) ). This arises from the discrete nature of the eigenvalues, which
is induced by the second order differential operator and the vanishing of
the eigenmodes at the positions of the singularities $\approx \pm y_{h}$.
The singularities are generated by the vanishing of the norm of the operator 
$\widehat{\alpha }$, which makes ambiguous the assumption of propagating
wave character, $\partial _{t}=-v_{d}\partial _{y}$. The large-$k$ part
mainly reflects the structure of the small-scale shape perturbation of the
soliton, comming from $\beta $-related terms. Fig.6 is an $\left( k,\omega
\right) $ spectrum obtained from $\omega -ku=0$ and repeating the
calculations for various soliton velocities $u_{\max }>u>v_{d}$. Although we
cannot afford high $u_{\max }$ since the expressions of $t_{1,2,3}\left(
y\right) $ depend on the assumption $u\gtrsim v_{d}$ , we remark local peaks
in contrast to the ``pure soliton'' result of Ref.\cite{Pf25-82-1838}.

For simplicity we have assumed a single soliton. However the calculation can
be readily extended to the multi-soliton case, considering instead of (\ref
{fijs}) and (\ref{hijs}) sums over many individual soliton solutions with
different velocities and positions of the centres. These sums replace the
functions $\varphi _{Js}$ and $\chi _{Js}$ in the expressions of the
operators $\widehat{\alpha }$, $\widehat{\beta }$ and $\widehat{\gamma }$.
If the velocities are all greater but not too different of $v_{d}$ the
change of variables to the referential moving with $v_{d}$ (described in the
paragraph below Eq.(\ref{oper12}) ) will leave a very slow time variation
which eventually may be treated perturbatively. Many solitons will also
generate many singularities arising from the vanishing of the function $h$,
and this will factorize the space of functions and correspondingly the
generating functional. It will become however possible to consider random
positions and random velocities and average them with distribution functions
for the Gibbs ensemble, like in \cite{Pf25-82-1838}. This is very simple
with the first term of (\ref{cor7}), which should be compared directly with
Ref.\cite{Pf25-82-1838}, but technically very difficult with the terms
involving functional derivatives of $A$ and/or $B$.

The first results suggests that the non-gaussianity at the plasma edge can
be explained by the presence of coherent structures. The contribution of
avalanches to the deviation from the gaussian statistics cannot be excluded
but, as shown for self-organized systems \cite{HwaKardar}, they have a
scaling which should be easy recognized, at least in frequency domain.

In conclusion we have developed an approach which allows us to calculate the
statistical properties of a coherent structure in a turbulent background.
Compared to the standard renormalization, this approach is at the opposite
limit in what concerns the relation ``coherent structure / wave
turbulence'', highlightning the coherent structure. However it offers
comparatively greater possibilities for the extension of this studies to the
more realistic problem of cascading wave turbulence mixed with rising and
decaying coherent structures.

{\bf Acknowledgments}. The authors are indebt to J.H. Misguich and R.
Balescu for many stimulating and enlightening discussions. F. S. and M.V.
gratefully acknowledge the support and hospitality of the {\it %
D\'{e}partement de Recherche sur la Fusion Control\'{e}e}, Cadarache, France.

This work has been partly supported by the NATO Linkage Grant CRG.LG 971484.

\bigskip

\appendix

\section{Explicit expressions for the functional derivatives}

We shall first concentrate on the derivatives of the two factors $A$ and $B$%
. 
\begin{eqnarray}
\frac{\delta B}{\delta J\left( y_{1}\right) } &=&\frac{\delta }{\delta
J\left( y_{1}\right) }\left[ \frac{\sigma /2}{\sin \left( \sigma /2\right) }%
\right] ^{1/2}  \label{cor8} \\
&=&\frac{1}{4}\left[ \frac{1}{\sin \left( \sigma /2\right) }+\frac{\sigma }{2%
}\frac{\cos \left( \sigma /2\right) }{\sin ^{2}\left( \sigma /2\right) }%
\right] \left[ \frac{\sigma /2}{\sin \left( \sigma /2\right) }\right]
^{-1/2}\left( \frac{\delta \sigma }{\delta J\left( y_{1}\right) }\right) 
\nonumber
\end{eqnarray}
and 
\begin{eqnarray}
&&\frac{\delta ^{2}B}{\delta J\left( y_{2}\right) \delta J\left(
y_{1}\right) }=\frac{\delta ^{2}}{\delta J\left( y_{2}\right) \delta J\left(
y_{1}\right) }\left[ \frac{\sigma /2}{\sin \left( \sigma /2\right) }\right]
^{1/2}  \label{cor9} \\
&=&\left\{ -\frac{1}{8}\left[ \frac{\sigma /2}{\sin \left( \sigma /2\right) }%
\right] ^{1/2}\frac{1+\cos ^{2}\left( \sigma /2\right) }{\sin ^{2}\left(
\sigma /2\right) }\right.  \nonumber \\
&&\left. -\frac{1}{16}\left[ \frac{\sigma /2}{\sin \left( \sigma /2\right) }%
\right] ^{-3/2}\left[ \frac{1}{\sin \left( \sigma /2\right) }+\frac{\sigma }{%
2}\frac{\cos \left( \sigma /2\right) }{\sin ^{2}\left( \sigma /2\right) }%
\right] ^{2}\right\} \,\left( \frac{\delta \sigma }{\delta J\left(
y_{2}\right) }\right) \left( \frac{\delta \sigma }{\delta J\left(
y_{1}\right) }\right)  \nonumber \\
&&+\frac{1}{4}\left[ \frac{\sigma /2}{\sin \left( \sigma /2\right) }\right]
^{-1/2}\left[ \frac{1}{\sin \left( \sigma /2\right) }+\frac{\sigma }{2}\frac{%
\cos \left( \sigma /2\right) }{\sin ^{2}\left( \sigma /2\right) }\right]
\;\left( \frac{\delta ^{2}\sigma }{\delta J\left( y_{2}\right) \delta
J\left( y_{1}\right) }\right)  \nonumber
\end{eqnarray}

For the exterior domains, 
\begin{equation}
\sigma =\sigma _0+\widetilde{\sigma }_{J1}+\widetilde{\sigma }_{J2}
\label{cor10}
\end{equation}
with 
\begin{equation}
\sigma _0=\frac 12\int_{-L/2}^{-y_h}dy^{\prime }\left[ -\left( \frac{%
\partial \varphi _s}{\partial y}\right) \frac{\frac{2c}h-1}{\left(
h^2-\varphi _s^2\right) ^{1/2}}\right]  \label{cor11}
\end{equation}
\begin{equation}
\widetilde{\sigma }_{J1}=\frac 12\int_{-L/2}^{-y_h}dy^{\prime }\left[ \left( 
\frac{\partial \varphi _s}{\partial y}\right) \frac 1{h\left( h^2-\varphi
_s^2\right) ^{1/2}}\left( 2-\frac{\varphi _s\left( 2c-h\right) }{h^2-\varphi
_s^2}\right) \widetilde{\chi }_J^{ext}\right]  \label{cor12}
\end{equation}
\begin{equation}
\widetilde{\sigma }_{J2}=\frac 12\int_{-L/2}^{-y_h}dy^{\prime }\left[ -\frac
1{\left( h^2-\varphi _s^2\right) ^{1/2}}\left( \frac{\partial \widetilde{%
\chi }_J^{ext}}{\partial y}\right) \right]  \label{cor13}
\end{equation}
We have the following connected expressions: 
\begin{equation}
\frac{\delta \sigma }{\delta J\left( y_1\right) }=\frac{\delta \widetilde{%
\sigma }_{J1}}{\delta J\left( y_1\right) }+\frac{\delta \widetilde{\sigma }%
_{J2}}{\delta J\left( y_1\right) }  \label{cor14}
\end{equation}
\begin{equation}
\frac{\delta \widetilde{\sigma }_{J1}}{\delta J\left( y_1\right) }=\frac
12\int_{-L/2}^{-y_h}dy^{\prime }\left( \frac{\partial \varphi _s}{\partial y}%
\right) \frac 1{h\left( h^2-\varphi _s^2\right) ^{1/2}}\left( 2-\frac{%
\varphi _s\left( 2c-h\right) }{h^2-\varphi _s^2}\right) \left( \frac{\delta 
\widetilde{\chi }_J^{ext}}{\delta J\left( y_1\right) }\right)  \label{cor15}
\end{equation}
\begin{equation}
\frac{\delta \widetilde{\sigma }_{J2}}{\delta J\left( y_1\right) }=\frac
12\int_{-L/2}^{-y_h}dy^{\prime }\frac{\left( -1\right) }{\left( h^2-\varphi
_s^2\right) ^{1/2}}\frac \delta {\delta J\left( y_1\right) }\left( \frac{%
\partial \widetilde{\chi }_J^{ext}}{\partial y}\right)  \label{cor16}
\end{equation}
and: 
\begin{equation}
\frac{\delta ^2\sigma }{\delta J\left( y_2\right) \delta J\left( y_1\right) }%
=\frac{\delta ^2\widetilde{\sigma }_{J1}}{\delta J\left( y_2\right) \delta
J\left( y_1\right) }+\frac{\delta ^2\widetilde{\sigma }_{J2}}{\delta J\left(
y_2\right) \delta J\left( y_1\right) }  \label{cor17}
\end{equation}
\begin{equation}
\frac{\delta ^2\widetilde{\sigma }_{J1}}{\delta J\left( y_2\right) \delta
J\left( y_1\right) }=\frac 12\int_{-L/2}^{-y_h}dy^{\prime }\left( \frac{%
\partial \varphi _s}{\partial y}\right) \frac 1{h\left( h^2-\varphi
_s^2\right) ^{1/2}}\left( 2-\frac{\varphi _s\left( 2c-h\right) }{h^2-\varphi
_s^2}\right) \left( \frac{\delta ^2\widetilde{\chi }_J^{ext}}{\delta J\left(
y_2\right) \delta J\left( y_1\right) }\right)  \label{cor18}
\end{equation}
\begin{equation}
\frac{\delta ^2\widetilde{\sigma }_{J2}}{\delta J\left( y_2\right) \delta
J\left( y_1\right) }=\frac 12\int_{-L/2}^{-y_h}dy^{\prime }\frac{\left(
-1\right) }{\left( h^2-\varphi _s^2\right) ^{1/2}}\;\frac{\delta ^2}{\delta
J\left( y_2\right) \delta J\left( y_1\right) }\left( \frac{\partial 
\widetilde{\chi }_J^{ext}}{\partial y}\right)  \label{cor19}
\end{equation}

For the ``interior'' region, the derivatives of $A$, (which are
strightforward) will require the calculation of the derivatives of $\beta $. 
\[
\beta =\frac 12\int_0^{y_h}dy\frac{-\frac 1{v_d}\left( \frac{\partial
\varphi _s}{\partial y}\right) \frac{2c-h}{h^3}+\frac 2{v_d}\frac
1{h^3}\left( \frac{\partial \varphi _s}{\partial y}\right) \widetilde{\chi }%
_J^{int}-\frac 1{v_d}\frac 1{h^2}\frac{d\widetilde{\chi }_J^{int}}{dy}}{%
\left( \frac 1{v_d^2}\frac{\varphi _s^2-h^2}{h^4}\right) ^{1/2}\left( 1-%
\frac{2\varphi _s}{\varphi _s^2-h^2}\widetilde{\chi }_J^{int}\right) ^{1/2}} 
\]
The function $\widetilde{\chi }_J^{int}$ and its derivative are present in
the expression of $\beta $: 
\[
\beta =\beta _0+\widetilde{\beta }_{J1}+\widetilde{\beta }_{J2} 
\]
\[
\beta _0=\frac 12\int_0^{y_h}dy\left[ -\left( \frac{\partial \varphi _s}{%
\partial y}\right) \frac{2c-h}{h\left( \varphi _s^2-h^2\right) ^{1/2}}\right]
\]
\[
\widetilde{\beta }_{J1}=\frac 12\int_0^{y_h}dy\left[ \left( \frac{\partial
\varphi _s}{\partial y}\right) \frac 1{h\left( \varphi _s^2-h^2\right)
^{1/2}}\left( 2-\frac{\varphi _s\left( 2c-h\right) }{\varphi _s^2-h^2}%
\right) \widetilde{\chi }_J^{int}\right] 
\]
\[
\widetilde{\beta }_{J2}=\frac 12\int_0^{y_h}dy\left[ -\frac 1{\left( \varphi
_s^2-h^2\right) ^{1/2}}\left( \frac{d\widetilde{\chi }_J^{int}}{dy}\right) %
\right] 
\]
and the derivatives at $J$ are easily calculated, as for $\sigma $.

The formulas above need to specify the expression of the functions $%
\widetilde{\chi }_{J}^{ext}$ , $\frac{\partial \widetilde{\chi }_{J}^{ext}}{%
\partial y}$ and of their functional derivatives. We use the results of the
analysis carried out by Karpman.

\newpage \label{biblio}

\smallskip

\newpage

\begin{center}
{\bf Figure Captions}
\end{center}

{\bf Fig.1} The form \smallskip of the soliton $\varphi _s(y) $
for the velocity $u=1.725\,v_d$.

{\bf Fig.2} The function $t_1\left( y\right) $ for $u=1.725\,v_d$.

{\bf Fig.3} The function $t_2\left( y\right) $ of the Eq.(\ref{wsec}) for
the same $u$.

{\bf Fig.4} The perturbation to the correlation in physical space.

{\bf Fig.5} Contour plot of the spectrum of the vortex perturbed by the
turbulent drift waves.

{\bf Fig.6 }The contour plot of the frequency-wavenumber spectrum, with $%
\omega -ku=0$.

\newpage

\begin{figure}
\centerline{
 \psfig{file=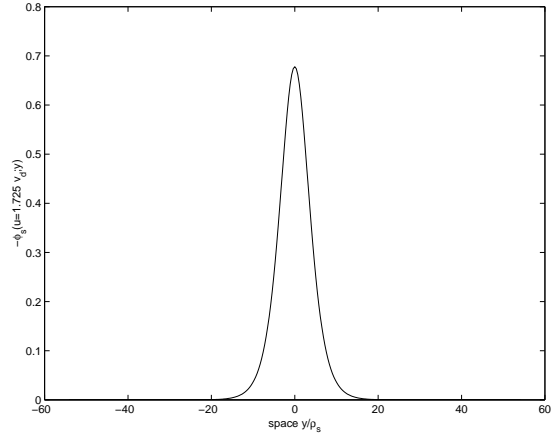,width=0.4\textwidth}}
\caption{{\bf Fig.1} Variation of the form \smallskip of the soliton $\varphi _s(y) $
with the velocity, $u$.
}
\end{figure}

\begin{figure}
\centerline{
 \psfig{file=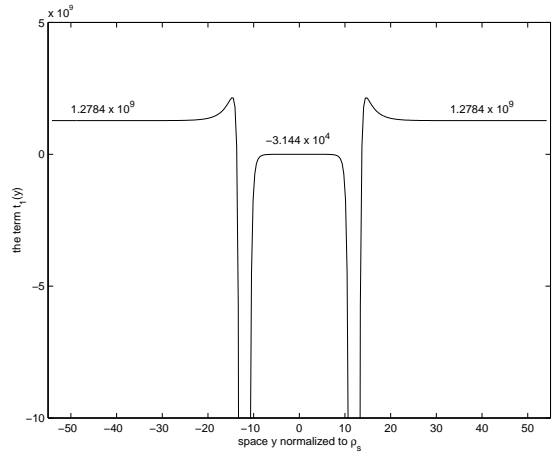,width=0.4\textwidth}}
\caption{{\bf Fig.2} The function $t_1\left( y\right) $ for a particular soliton
velocity, $u=1.725\,v_d$.
}
\end{figure}

\begin{figure}
\centerline{
 \psfig{file=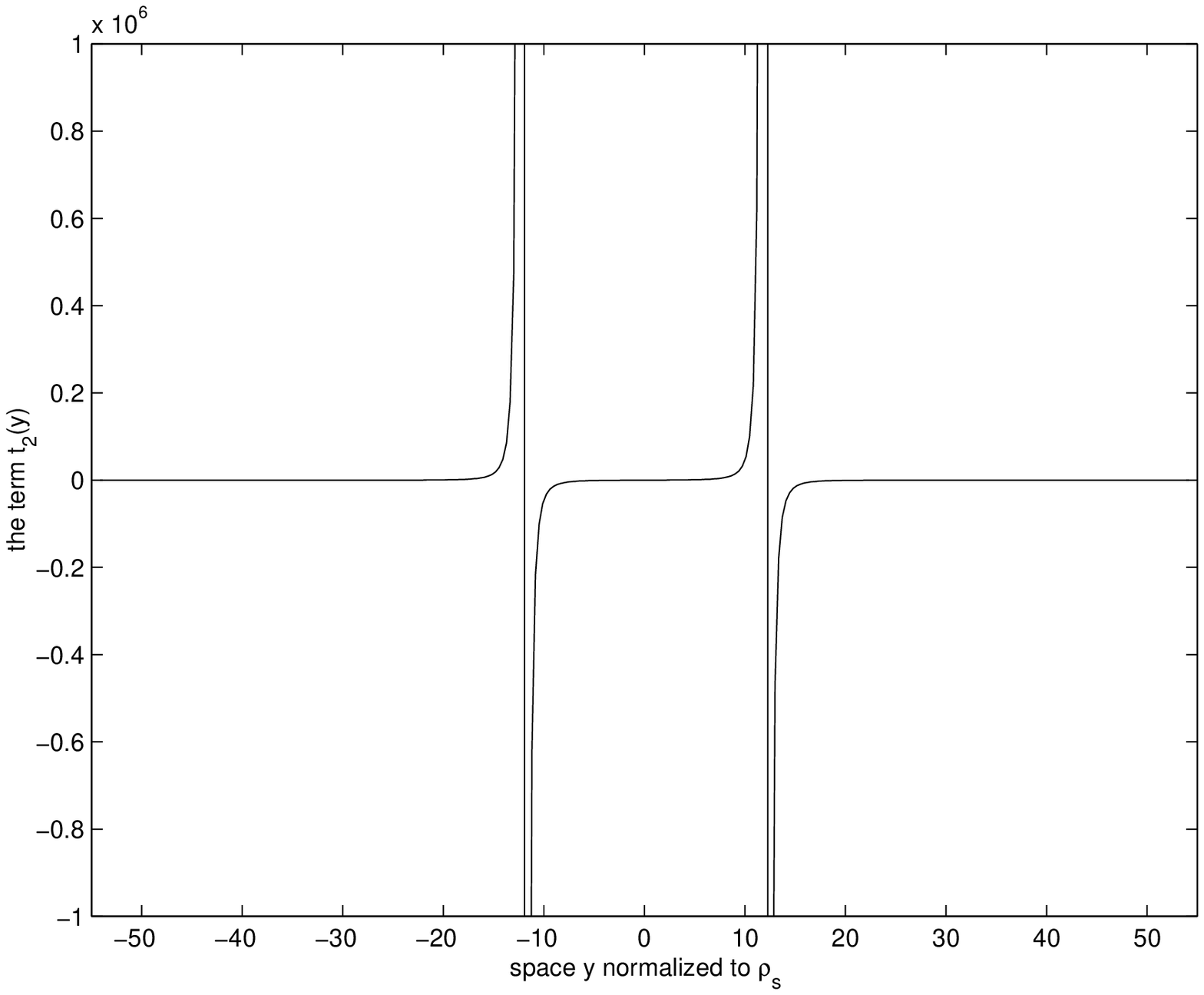,width=0.4\textwidth}}
\caption{{\bf Fig.3} The function $t_2\left( y\right) $ of the Eq.(\ref{wsec}) for
the same $u$.
}
\end{figure}

\begin{figure}
\centerline{
 \psfig{file=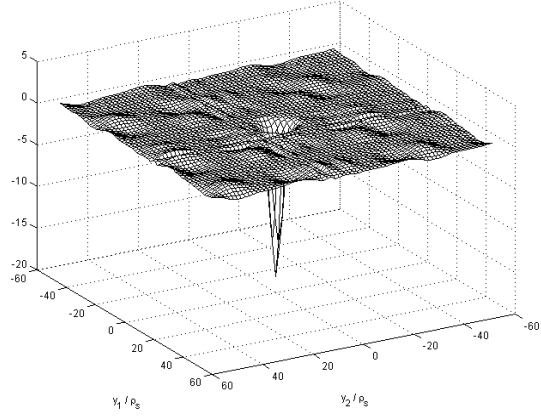,width=0.4\textwidth}}
\caption{{\bf Fig.4} The perturbation to the correlation in physical space.}
\end{figure}

\begin{figure}
\centerline{
 \psfig{file=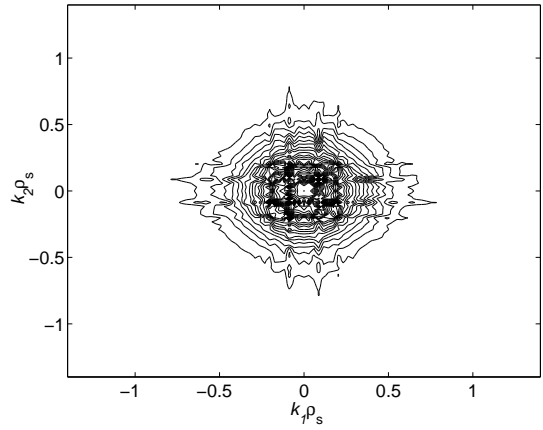,width=0.4\textwidth}}
\caption{{\bf Fig.5} Contour plot of the spectrum of the vortex perturbed by the
turbulent drift waves.}
\end{figure}

\begin{figure}
\centerline{
 \psfig{file=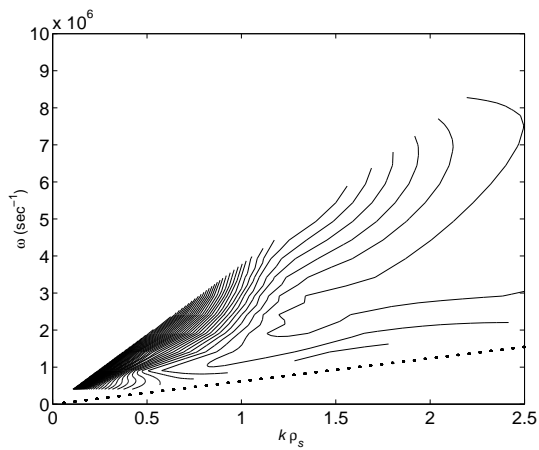,width=0.4\textwidth}}
\caption{{\bf Fig.6 }The contour plot of the frequency-wavenumber spectrum, with $\omega -ku=0$.}
\end{figure}

\end{document}